\documentclass[12pt]{iopart}
\usepackage{iopams}
\usepackage[pdftex]{graphicx}
\usepackage{xcolor,soul}
\begin{document}

\title[Mass--energy connection without special relativity]{Mass--energy
  connection without special relativity}

\author{Germano D'Abramo}

\address{Ministero dell'Istruzione, dell'Universit\`a e della Ricerca,
  I-00041, Albano Laziale, RM, Italy}
\ead{germano.dabramo@gmail.com}
\vspace{10pt}
\begin{indented}
\item[]September, 2020
\item[]  
\end{indented}

\begin{abstract}
  In 1905, Einstein carried out his first derivation of the mass--energy
  equivalence by studying in different reference frames the energy balance of
  a body emitting electromagnetic radiation and assuming special relativity
  as a prerequisite. In this paper, we prove that a general mass--energy
  relationship can be derived solely from very basic assumptions, which are
  the same made in Einstein's first derivation but completely neglecting
  special relativity. The general mass--energy relationship turns to a
  mass--energy equivalence when is applied to the case of a body emitting
  energy in the form of electromagnetic waves.
  Our main result is that if the core logic behind Einstein's approach
  is sound, then the essence of the mass--energy equivalence can be derived
  without special relativity. We believe that our heuristic approach, although
  not capable of giving the exact mathematical formula for the mass-energy
  equivalence, may represent a useful addition to the general discussion on
  the matter at the graduate level. Our finding suggests that the connection
  between mass and energy is at a deeper level and comes before any full-fledged
  physical theory.
      
\end{abstract}

%
\vspace{2pc}

\submitto{\EJP}
%
%
%

\section{Introduction}
\label{se1}

Mass--energy equivalence, known in the form of the celebrated equation
$E=mc^2$, was derived by Einstein  for the first time in a three-page paper
published at the end of 1905~\cite{e05b}. Einstein carried out his derivation
by studying in different reference frames the energy balance of a body emitting
electromagnetic radiation in two equal but oppositely directed amounts (thus,
no change in the emitter velocity due to recoil). According to special
relativity~\cite{e05}, the total energy of a plane light wave increases when
is observed from a reference frame in uniform motion relative to the emitter's
rest frame. Einstein ascribed this increase to the fact that in the moving
reference frame also the total energy of the emitter, where the radiation
energy comes from, has increased: when the emitter is observed from the moving
reference frame, its kinetic energy must also be added to its internal (proper)
energy to get its total energy. Then, Einstein managed to derive that the
increase of the emitted energy seen from the moving frame comes from a
reduction in kinetic energy of the emitter after the emission. Since, for
symmetry reasons, the velocity of the emitter does not change after the
emission, Einstein concluded that the mass of the emitter must change by
partially turning into radiation energy.

The correctness of this derivation was first criticized by Planck in
1907~\cite{p07}. He contended that it is valid ``under the assumption
permissible only as a first approximation that the total energy of a body is
composed additively of its kinetic energy and its energy referred to a system
in which it is at rest''~\cite{j61}. Further criticism was later advanced by
Ives in 1952~\cite{i52} and Jammer in 1961~\cite{j61}: they asserted that
Einstein's derivation was but the result of a {\em petitio principii}. Several
other authors (e.g.~G.~Holton, H.~Arzeli\'es and A.I.~Miller, to name a few)
agreed with Ives and Jammer criticism. Recently, however, Stachel and
Torretti~\cite{st82} analyzed Ives's analysis and concluded that the logic
behind Einstein's derivation is sound.
In particular, they presented a proof from first principles of the assumption
criticized by Planck. We shall return briefly to their analysis later on.
In more recent times, Ohanian~\cite{o08,o08b} agreed with Stachel and Torretti's
criticism of Ives, though he argued that Einstein's derivation was wrong mainly
``because he assumed that the rest-mass change he found when using a
non-relativistic, Newtonian approximation for the internal motions of an
extended system would be equally valid for relativistic motions''.

\begin{figure}[t]
\begin{center}
\includegraphics[width=12cm]{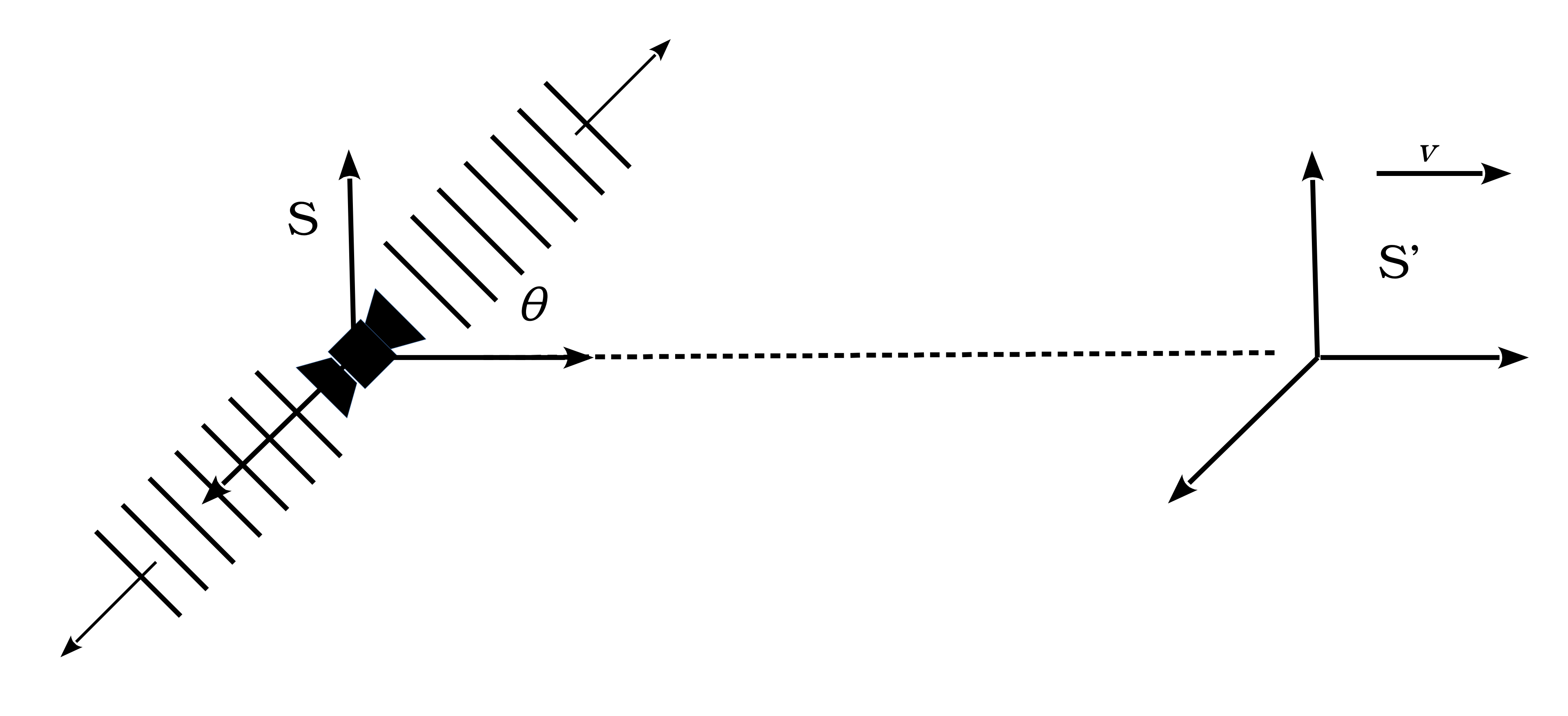}
\end{center}
\caption{Sketch of the light emission process described in
  Einstein's paper~\cite{e05b}.}
\label{fig1}
\end{figure}

For the sake of completeness, let us review Einstein's first derivation in
more detail. Einstein considered a body, at rest in an inertial frame $S$, that
emits electromagnetic radiation of total energy L in two equal but oppositely
directed amounts. He then considered the same emission process as seen from
another inertial frame $S'$, that of an observer moving in uniform parallel
translation with respect to the system $S$ and having its origin of coordinates
in motion along the $x$-axis with velocity $v$ (Fig.~\ref{fig1}).

Let there be a stationary body in the system $S$, and let its energy referred
to the system $S$ be E$_0$. Let the energy of the body relative to the system
$S'$ moving as above with velocity $v$, be $\textrm{E}'_0$.

Let this body send out, in a direction making an angle $\theta$ with the
$x$-axis, plane waves of light of energy $\frac{1}{2}$L measured relatively
to $S$, and simultaneously an equal quantity of plane waves in the
opposite direction, for a total emitted energy equal to L (see
Fig.~\ref{fig1}). Meanwhile, the body remains at rest in $S$.

Einstein showed that if the radiation is measured in $S'$, then it possesses
a total energy L$'$ that is equal to

\begin{equation}
\textrm{L}'=\frac{\textrm{L}}{\sqrt{1-\frac{v^2}{c^2}}},
\label{eq1}
\end{equation}
where $c$ is the velocity of light. This relation is the result established
by using the law for the transformation of the energy of a plane light wave
from one inertial frame to the other, derived in the first paper on the special
relativity~\cite{e05}.

If we call the energy of the body after the emission of the plane light waves
E$_1$ or $\textrm{E}'_1$ respectively, measured relatively to the system $S$ or
$S'$ respectively, then by making use of eq.~(\ref{eq1}) we have

\begin{equation}
\begin{array}{lr}
  \textrm{E}_0 = \textrm{E}_1+\textrm{L},\\
         \textrm{E}'_0 = \textrm{E}'_1
         +\frac{\textrm{L}}{\sqrt{1-\frac{v^2}{c^2}}}.
\end{array}
\label{eq2}
\end{equation}

By subtraction, Einstein obtained the following relation

\begin{equation}
(\textrm{E}'_0-\textrm{E}_0)-(\textrm{E}'_1-\textrm{E}_1)=
         L\left\{\frac{1}{\sqrt{1-\frac{v^2}{c^2}}}-1\right\}.
\label{eq3}
\end{equation}

According to Einstein's reasoning, the two differences of the form
$\textrm{E}'-\textrm{E}$ in eq.~(\ref{eq3}) have the following simple physical
meaning. E$'$ and E are the energy values of the same body referred to two
reference frames that are in motion relatively to each other, the body being at
rest in $S$. Thus, the difference $\textrm{E}'-\textrm{E}$ can differ from the
kinetic energy K of the body, with respect to the system $S'$, only by an
additive constant C, which depends on the choice of the arbitrary additive
constants of the energies E$'$ and E and does not change during the emission of
light. Without loss of generality, this constant can be taken equal to zero,
and the difference can be written simply as $\textrm{E}'-\textrm{E}=\textrm{K}$.
This assumption drew the attention of most of the following literature
on the first mass--energy equivalence derivation and generated some
controversy on its validity. A careful discussion of this aspect is given
in~\cite{st82}. In the same paper, the authors give a formal derivation of
Einstein's assumption from first principles, and their approach is presented
as general. As already mentioned, according to these authors, Einstein's
assumption turns out to be logically sound. In any case, the validity of what
we shall present in Section~\ref{se2} also relies on the acceptance of the
validity of this assumption.

From eq.~(\ref{eq3}) we have

\begin{equation}
  \textrm{K}_0-\textrm{K}_1=
  L\left\{\frac{1}{\sqrt{1-\frac{v^2}{c^2}}}-1\right\}.
\label{eq4}
\end{equation}

What equation~(\ref{eq4}) tells us is that the kinetic energy of the body with
respect to $S'$ diminishes as a result of the emission of the plane light
waves, and the amount of diminution is independent of the properties of the
body. Moreover, like the kinetic energy, it depends on the relative velocity
$v$. Neglecting quantities of the fourth and higher orders in $v/c$,
eq.~(\ref{eq4}) becomes

\begin{equation}
  \textrm{K}_0-\textrm{K}_1=
   \frac{1}{2}\left[\frac{\textrm{L}}{c^2}\right]v^2.
\label{eq4b}
\end{equation}

From eq.~(\ref{eq4b}), Einstein's mass--energy equivalence directly follows:
if a body gives off the energy L (in the form of radiation), its mass
diminishes by $\frac{\textrm{L}}{c^2}$.

The rest of this paper is organized as follows. In Section~\ref{se2}, we prove
that it is possible to derive a general mass--energy relationship by
following the logic behind Einstein's original derivation and by applying the
same fundamental assumptions but neglecting special relativity. Within the
sphere of validity of these basic assumptions, the general mass--energy
relationship would still be true even if special relativity would turn out to
be false. We also notice that the general mass--energy relationship turns to a
mass--energy equivalence when is applied to the case of a body emitting energy
in the form of electromagnetic waves: this is the crucial step in Einstein's
first derivation, and special relativity turns out to have no fundamental role
in the realness of the equivalence.
We shall show that mass--energy equivalence, although with a different
mathematical equation, could have been derived even within Maxwell's theory of
light (pre-Lorentz, classical ether theory).

In the concluding section, we summarize our findings and remark why they
represent a useful addition to the general discussion on the matter.

\section{The general mass--energy relationship}
\label{se2}

It is possible to heuristically derive a general mass--energy relationship by
applying the core logic behind Einstein's original derivation but without
special relativity. We only use few and very basic initial assumptions which
are the same made in Einstein's derivation, exception made for the peculiar
principles of special relativity.

Consider a body stationary in an inertial frame $S$ that emits a total amount
of energy equal to L. The energy can be emitted in any imaginable form but,
like in Einstein's derivation, always in equal amounts in opposite directions
to maintain a symmetry of emission that intuitively ensures the motionlessness
of the body during the process. The equation of the energy balance in $S$ is
then $\textrm{E}_0 = \textrm{E}_1+\textrm{L}$, where $\textrm{E}_0$ and
$\textrm{E}_1$ are the total energies of the body respectively before and after
the emission referred to the system $S$.

If the same emission process is seen from an inertial reference frame $S'$
moving in uniform parallel translation with respect to the system $S$ and
having its origin of coordinates in motion along the $x$-axis with velocity
$v$, then it is reasonable to expect that the observed
{\em total}\footnote{We invite the reader to pay attention to the
use of the word `total' here. We know from experience (e.g.~with sound
waves, light waves, etc.) that the carried energy is `perceived' as higher or
lower according to the emission direction relative to the observer. However,
here we consider the {\em overall} energy emitted by the source, namely the
sum (integral) of the energy emitted in any direction. A corroboration of the
fact that we expect greater overall energy is given further in the text when
we calculate the energy of two light waves within Maxwell's theory of light,
eqs.~(\ref{eq17}) to~(\ref{eq22}).} emitted energy $\textrm{L}'$ is different
from L and greater than that.
This is what we heuristically expect in real life simply because the observer
is moving relative to the emitter, and some energy is added to what he sees
because of that motion. The equation of the energy balance in $S'$
is then $\textrm{E}'_0=\textrm{E}'_1 + \textrm{L}'$, where $\textrm{E}'_0$ and
$\textrm{E}'_1$ are the total energies of the body respectively before and
after the emission referred to the system $S'$. So far, we have used only the
principle of energy conservation in any inertial frame.

Without loss of generality, we can write the mathematical relation that
connects L$'$ and L as follows

\begin{equation}
  \textrm{L}'={\cal F}(\textrm{L}, v),
\label{eq6}
\end{equation}
where ${\cal F}$ is a suitable mathematical function. Since the origin of
reference frame $S'$ moves along the $x$-axis, the functional dependence of
eq.~(\ref{eq6}) on velocity is by construction on scalar velocity ${v}$.
Moreover, let L$'$ be directly proportional to L. If the body emits energy
equal to 2L, the energy observed in $S'$ must be equal to 2L$'$. Indeed, this
seems a reasonable assumption: the body emitting energy 2L can, in theory, be
composed of two distinct bodies emitting energy L each. Since in this second
case the observer in $S'$ sees a total energy of 2L$'$ (L$'$ for each body),
this must be also the case when we have a single body emitting energy equal
to 2L. Thus, equation~(\ref{eq6}) becomes

\begin{equation}
  \textrm{L}'=\textrm{L}f(v).
\label{eq7}
\end{equation}

In order to determine the approximate mathematical form of the dimensionless
function $f(v)$, consider the Maclaurin expansion of $f(v)$ up to $O(v^3)$

\begin{equation}
  f(v)=\alpha+\beta v + \delta v^2 + O(v^3),
\label{eq8}
\end{equation}
where $\alpha$, $\beta$, and $\delta$ are numerical coefficients.

Since $f(0)=1$, $\alpha$ must be equal to 1. Furthermore, we must have that
$f(-v)= f(v)$ since, for symmetry reasons\footnote{Whatever is the direction
$\theta$ along which the energies L/2 are emitted (see Fig.~\ref{fig1}), the
case in which we observe $S$ and move in translational motion towards the
positive $x$-axis ($+v$) is, as a whole, physically equivalent to the case in
which we observe $S$ and move in translational motion towards the negative
$x$-axis ($-v$), provided that the whole setting is flipped over the $x$-axis.
The amount of energy $\textrm{L}'$ cannot change because of these symmetry
(abstract) operations.}, the {\em overall} energy L$'$ observed by an observer
in $S'$ does not depend upon the arbitrary direction (towards the positive or
the negative $x$-axis) of the velocity of $S'$ and thus $\beta =0$. 
Since $f(-v)= f(v)$, function $f(v)$ must be even, and all other
terms with odd powers must be absent. Therefore,

\begin{equation}
  f(v)= 1 + \delta v^2 + O(v^4),
\label{eq9}
\end{equation}
with constant $\delta$ having the physical units of an inverse square
velocity. This velocity is the `characteristic velocity' of the peculiar
emission process.

Thus, we arrive at

\begin{equation}
\textrm{L}'=\textrm{L} (1+\delta v^2+O(v^4)).
\label{eq10}
\end{equation}
Within the sphere of validity of the previous assumptions, equation~(\ref{eq10})
is very general and can be applied to all kinds of energy emission mechanisms.
As a matter of fact, its derivation is completely independent of the
energy emission process at play, exception made for the numerical value of the
constant $\delta$. 

Now, the energy balance equations become

\begin{equation}
\begin{array}{lr}
  \textrm{E}_0 = \textrm{E}_1+\textrm{L},\\
  \textrm{E}'_0 = \textrm{E}'_1 +\textrm{L} (1+\delta v^2+O(v^4)).
\end{array}
\label{eq11}
\end{equation}

Like Einstein in his 1905 paper, we subtract the first equation from the
second

\begin{equation}
(\textrm{E}'_0-\textrm{E}_0)-(\textrm{E}'_1-\textrm{E}_1)=
  \textrm{L}(\delta v^2+O(v^4)),
\label{eq12}
\end{equation}
and with Einstein's assumption $\textrm{E}' - \textrm{E}=\textrm{K}$ we
obtain

\begin{equation}
  \textrm{K}_0-\textrm{K}_1=
       L(\delta v^2+O(v^4)).
\label{eq13}
\end{equation}

If, like in~\cite{st82}, we define the inertial mass for a body in
translational motion (in keeping with the requirement that special relativistic
dynamics have a Newtonian limit as $v\to 0$) by

\begin{equation}
  m=\lim_{v\to 0}\frac{K}{v^2/2},
  \label{eq14}
\end{equation}
then from eq.~(\ref{eq13}) it follows

\begin{equation}
  -\Delta m= m_0-m_1=\lim_{v\to 0}\frac{(K_0-K_1)}{v^2/2}=
  \lim_{v\to 0}\frac{L(\delta v^2 + O(v^4))}{v^2/2}=2\delta\textrm{L}.
  \label{eq15}
\end{equation}

In short,

\begin{equation}
  -\Delta m=2\delta\textrm{L},
  \label{eq16}
\end{equation}
and this is an exact, not an approximate result. If a body gives off the energy
L, its mass diminishes by $2\delta\textrm{L}$.

Notice that eq.~(\ref{eq16}) is not a mass--energy equivalence {\em per se}.
If we apply eq.~(\ref{eq16}) to a body releasing two projectiles of mass $m$ in
opposite directions with non-relativistic velocity $v_0$ (relative to the
parent body), then it is possible to prove that $\delta=1/v_0^2$. Since
$\textrm{L}=2\cdot\frac{1}{2}mv_0^2$ (the emitted energy, in this case, is only
kinetic), then $-\Delta m=2m$. Namely, equation~(\ref{eq16}) gives simply the
change of mass of the parent body due to the loss of two projectiles of mass
$m$ each. Thus, in this case, eq.~(\ref{eq16}) does not give any mass--energy
equivalence.

On the other hand, if we apply  eq.~(\ref{eq16}) to the emission of energy in
the form of electromagnetic waves, we obtain a mass--energy equivalence:
{\em radiation energy comes from mass reduction, and thus mass transforms into
radiation energy}.
Special relativity is not essential for the derivation of this
mass--energy equivalence: special relativity comes into play only in the
numerical value of the constant $\delta$. The constant $\delta$ has the
physical units of an inverse square velocity, and in the case of
electromagnetic phenomena, it must be heuristically proportional to $1/c^2$.
In the case of Einstein's original derivation, we have that $\delta=1/2c^2$.

In order to emphasize the implications of the derived general mass--energy
relationship, consider that even within Maxwell's theory of light (and thus,
no special relativity), one could have already come to mass--energy
equivalence, albeit in the different form $E =\frac{1}{2}mc^2$.

Within Maxwell's theory of light (pre-Lorentz, classical ether theory),
we have that $\delta=1/c^2$.
The total energy density associated with an electromagnetic wave is

\begin{equation}
u=\frac{1}{2}\epsilon_0E^2+\frac{1}{2}\frac{B^2}{\mu_0}=\epsilon_0E^2,
\label{eq17}
\end{equation}
where $\epsilon_0$ and $\mu_0$ are respectively the permittivity
and the permeability of free space, and $E$ and $B$ denote the
electric and magnetic fields of the wave. The last equality in
eq.~(\ref{eq17}) holds because, for electromagnetic waves, we also have that
$E=cB$ ($c=\frac{1}{\sqrt{\epsilon_0\mu_0}}$).
  
\begin{figure}[t]
\begin{center}
\includegraphics[width=9cm]{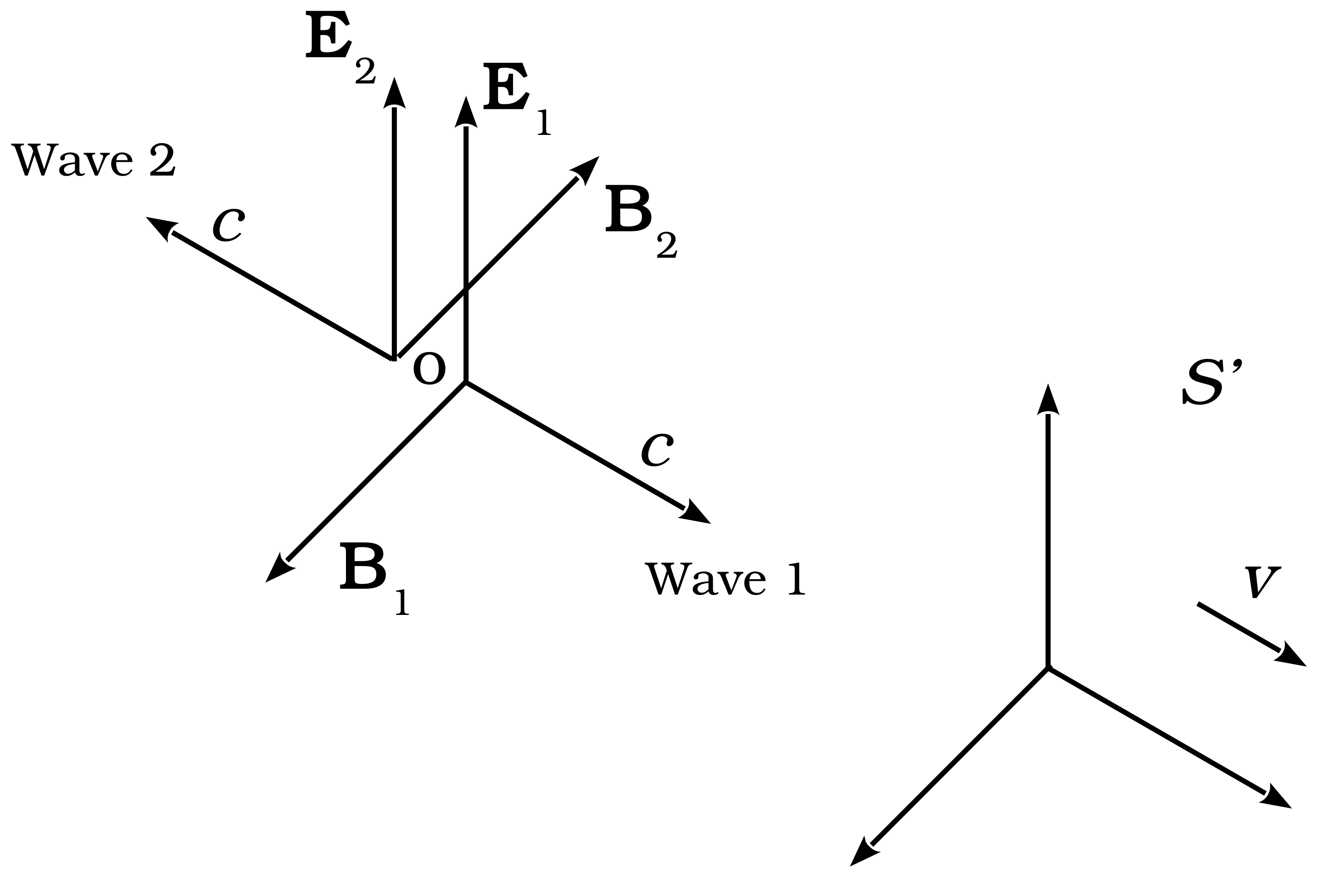}
\end{center}
\caption{Emission of light waves 1 and 2.}
\label{fig3}
\end{figure}

Now, consider two plane waves of light, 1 and 2, emitted in opposite
directions from the origin of the rest frame $S$ along the $x$-axis, as shown
in Fig.~\ref{fig3}. Since we are working within Maxwell's theory of light,
frame $S$ shall also be considered as the reference frame of the ether.
Consider further a reference frame $S'$ moving away from the origin of $S$
with velocity $v$ in the direction of the positive $x$-axis (in the
approximation $v\ll c$).
In the present context, we cannot use the Lorentz transformations for the
electromagnetic field to derive the electric field ${\bf E'}$ measured in the
reference frame $S'$. Nonetheless, it is possible to obtain a suitable
transformation law that applies to this specific case via the Lorentz force
${\bf F}=q({\bf E}+{\bf v}\times {\bf B})$ felt by a test charge $q$
stationary in $S'$, namely

\begin{equation}
{\bf E'}=\frac{{\bf F}}{q}={\bf E}+{\bf v}\times {\bf B}.
\label{eq18}
\end{equation}

See also reference~\cite{jac}, where the same result is obtained by applying
Faraday's law in the approximation of reference frames moving at speeds small
compared to the speed of light.

According to the above transformation, the components $E'_{1\parallel}$ and
$E'_{1\perp}$ of the electric field of wave~1 in the reference frame $S'$
are ($\parallel$ and $\perp$ are referred to the plane $(c,{\bf B})$, see
Fig.~\ref{fig3})

\begin{equation}
\begin{array}{lr}
    E'_{1\parallel} = E_{1\parallel}=0,\\
    E'_{1\perp} = E_{1\perp}+({\bf v}\times {\bf B}_1)_{\perp}=E\left(1-
  \frac{v}{c}\right),
\end{array}
\label{eq19}  
\end{equation}
since $E = E_{1\perp}$ and $B=E/c$.

The energy density $u_1'$ of wave~1 measured from $S'$ is then

\begin{equation}
  u'_1=\epsilon_0E^2\left(1-\frac{v}{c}\right)^2=
  u\left(1-\frac{v}{c}\right)^2.
\label{eq20}  
\end{equation}

By applying the same procedure to wave~2, the energy density $u_2'$ is

\begin{equation}
  u'_2=\epsilon_0E^2\left(1+\frac{v}{c}\right)^2=
  u\left(1+\frac{v}{c}\right)^2.
\label{eq21}  
\end{equation}

In order to calculate the energy of the two plane waves of light, we now need
to multiply the energy densities by the volumes of the plane waves measured in
$S'$. After an interval of time $T$, wave~1 has traveled a distance $cT$
from the origin of reference frame $S$, and its volume $V$ is simply $V=AcT$,
where $A$ is the transverse area of the wave. For an observer in $S'$, the
volume is the same since, in Maxwell's theory of light, light propagates
at speed equal to $c$ only in the ether reference frame $S$, and no Lorentz
contraction comes into play.

If the total energy of the two plane waves of light in $S$ is
$\textrm{L}=2uV$, then the total energy measured in $S'$ is

\begin{equation}
  \begin{array}{lr}
  \textrm{L}'=u'_1V_1'+u'_2V'_2=u\left(1-\frac{v}{c}\right)^2V+
  u\left(1+\frac{v}{c}\right)^2V=\\
   \\
  \quad =2uV\left(1+\frac{v^2}{c^2}\right)=
  \textrm{L}\left(1+\frac{v^2}{c^2}\right),
  \end{array}
\label{eq22}  
\end{equation}
and thus $\delta=1/c^2$.

\section{Concluding remarks}
\label{se3}

We have shown that a general mass--energy connection can be heuristically
derived by applying the core logic behind Einstein's original derivation with
very basic assumptions but neglecting special relativity. Einstein's 1905
mass--energy equivalence is a special case of this general relationship: the
general mass--energy connection turns to a mass--energy equivalence when is
applied to the case of a body emitting energy in the form of electromagnetic
waves. Obviously, to obtain the exact mathematical equation for the
mass--energy equivalence, we still need special relativity. Moreover, the
validity of our result stands upon the acceptance of the validity and logical
consistency of the basic assumptions in Einstein's original derivation.
However, within these confines, our finding shows that the mass--energy
equivalence seems to originate at a deeper, fundamental level and from first
and general principles.
In 1946, Einstein proposed an elementary derivation of the mass--energy
equivalence that allegedly does not presume the formal machinery of special
relativity but uses only three previously known laws of physics~\cite{e46}:
(i) the law of the conservation of momentum; (ii) the expression for the
pressure of radiation; (iii) the well-known expression for the aberration of
light. Our approach, instead, suggests that the mass--energy equivalence
is almost inescapable, as happens with new laws of physics derived from
dimensional analysis, and comes before any full-fledged physical theory.

We acknowledge that our approach is not capable of giving the exact
mathematical formula for the mass-energy equivalence, and thus it is not a
rigorous derivation of that equivalence.  However, conceptually speaking, it
is as rigorous as the proof attempts made by Einstein itself~\cite{h11} since
it derives from the same core logic behind most of them.

\ack
The author is indebted to Dr.~Assunta Tataranni and Dr.~Gianpietro Summa for
key improvements to the manuscript. The author would also like to acknowledge
three anonymous reviewers who made several suggestions for revision that have
greatly improved this paper.

\end{document}